\documentclass{article}
\usepackage{arxiv}
\usepackage{cite}
\usepackage{amsmath,amssymb,amsfonts}
\usepackage{algorithmic}
\usepackage{graphicx}
\usepackage{textcomp}
\usepackage{tabularx}
\usepackage{tabularx}
\usepackage{hyperref}       
\usepackage{ieeetrantools}
\usepackage{booktabs}
\usepackage{multirow}
\usepackage{subcaption}
\usepackage{makecell}
\newcolumntype{R}{>{\raggedleft\arraybackslash}X}
\def\BibTeX{{\rm B\kern-.05em{\sc i\kern-.025em b}\kern-.08em
    T\kern-.1667em\lower.7ex\hbox{E}\kern-.125emX}}

\hypersetup{
	pdftitle={Mitigating 3D Prostate Biparametric MRI Data Scarcity through Domain Adaptation using Locally-Trained Latent Diffusion Models for Prostate Cancer Detection},
	pdfsubject={eess.IV},
	pdfauthor={Emerson P.~Grabke, Babak Taati, Masoom A.~Haider},
	pdfkeywords={Generative AI, Large Language Model Adapters, Medical Image Generation, Prostate MRI, Domain Adaptation},
}
\renewcommand{\shorttitle}{CCELLA++: Mitigating 3D Prostate bpMRI Data Scarcity}
\renewcommand{\undertitle}{This work has been submitted to the IEEE for possible publication. Copyright may be transferred without notice, after which this version may no longer be accessible.}
\title{Mitigating 3D Prostate Biparametric MRI Data Scarcity through Domain Adaptation using Locally-Trained Latent Diffusion Models for Prostate Cancer Detection}
\usepackage{authblk}

\author[1,2,3]{%
	Emerson P.~Grabke\thanks{\texttt{e.grabke@mail.utoronto.ca}\qquad Corresponding Author}%
}
\author[1,3,5,6]{%
Babak Taati\thanks{\texttt{babak.taati@uhn.ca}}%
}
\author[1,2,4]{%
	Masoom A.~Haider\thanks{\texttt{m.haider@utoronto.ca}}%
}
\affil[1]{Institute of Biomedical Engineering, University of Toronto}
\affil[2]{Lunenfeld-Tanenbaum Research Institute, Mount Sinai Hospital}
\affil[3]{KITE Research Institute, Toronto Rehabilitation Institute, University Health Network}
\affil[4]{Joint Department of Medical Imaging, University of Toronto, Princess Margaret Hospital, and Sinai Health systems}
\affil[5]{Department of Computer Science, University of Toronto}
\affil[6]{Faculty Affiliate of the Vector Institute, Toronto}

\begin{document}
\bstctlcite{IEEEexample:BSTcontrol}
\maketitle
\begin{abstract} 
Objective: Latent diffusion models (LDMs) could mitigate data scarcity challenges affecting machine learning development for medical image interpretation. The recent CCELLA LDM improved prostate cancer detection performance using synthetic MRI for classifier training but was limited to the axial T2-weighted (AxT2) sequence, did not investigate inter-institutional domain shift, and prioritized PI-RADS over histopathology outcomes. Methods: We propose CCELLA++, a novel LDM pipeline for simultaneous 3D biparametric prostate MRI (bpMRI) generation, including the AxT2, high b-value diffusion series (HighB) and apparent diffusion coefficient map (ADC), to overcome these limitations. We investigated source-free domain adaptation with classifiers pretrained on single institution real or LDM-generated synthetic data prior to fine-tuning on fractions of an out-of-distribution, external dataset. Results: CCELLA++ achieved comparable AxT2 Kernel Inception Distance to CCELLA (0.0128, 0.0131 respectively). CCELLA++ synthetic bpMRI pretraining outperformed real bpMRI in AP and AUC up to 12.5\% (n$\leq$166) external dataset volume (p\textless0.01 all), no pretraining in AUC up to 25\% external volume (n=332, p\textless0.05 all), and CCELLA AxT2-only pretraining in both data-scarce (n=83, p\textless0.001 AP and AUC) and full data (n=1329, p\textless0.05 AP and AUC) scenarios. Conclusion: CCELLA++ synthetic bpMRI can improve downstream classifier generalization and performance beyond real bpMRI or CCELLA-generated AxT2-only images. Future work should quantify medical image quality, balance bpMRI LDM training, and condition the LDM with additional information. Significance: CCELLA++ can generate synthetic bpMRI that outperforms real data for domain adaptation with data-scarce external institutions, advancing machine learning development for medical imaging. Our code is available at \url{https://github.com/grabkeem/CCELLA-plus-plus}
\end{abstract}


\section{Introduction}
\label{sec:introduction}
Labeled data scarcity is a common challenge impeding machine learning (ML) advancements within the medical image interpretation field. Causes for labeled data scarcity include limited institutional cohort size, data sharing challenges surrounding patient privacy and data governance, and the time and financial costs required for expert radiologist annotation~\cite{cheplyginaNotsosupervisedSurveySemisupervised2019,litjensSurveyDeepLearning2017,changDistributedDeepLearning2018}. This in-turn limits training data availability, which can negatively impact ML performance. Two notable strategies for mitigating data scarcity within medical imaging are domain adaptation--adapting a pretrained model to a new domain with limited data volume for the same task~\cite{kimTransferLearningMedical2022}--and generative ML--creating synthetic data to artificially increase ML training data volume~\cite{guoMAISIMedicalAI2025}.

Prostate cancer detection from biparametric MRI (bpMRI) is a medical pathology and imaging method that benefits well from ML data scarcity research. Prostate cancer is the second most common cancer type in men globally, with MRI significantly reducing biopsy rates when included during the prostate cancer diagnostic workup~\cite{ProstateCancerStatistics,schootsMagneticResonanceImaging2025}. However, prostate MRI interpretation accuracy exhibits inter-reader variability and depends on reader expertise~\cite{stabileFactorsInfluencingVariability2020}. While ML could assist with interpretation, prostate MRI has little publicly available domain-specific data, with the largest public dataset consisting of 1500 samples~\cite{sahaArtificialIntelligenceRadiologists2024}. This relative paucity is reflected within latent diffusion model (LDM) for medical MRI synthesis literature, where prostate MRI generation is investigated much less than higher data-availability domains such as brain MRI~\cite{fanSurveyEmergingApplications2024}. LDM studies within prostate MRI also tend to be unimodal~\cite{grabkeLeveragingClinicalText2025,wildeMedicalDiffusionBudget2024}, limited to 2D images~\cite{wildeMedicalDiffusionBudget2024,saeedBiparametricProstateMR2024}, and restricted to curated keyword or short phrase use rather than using the rich textual information present in full-text clinical radiology reports~\cite{saeedBiparametricProstateMR2024}. Clinical guidelines assert however that the axial T2-weighted (AxT2) sequence, high b-value diffusion-weighted image (HighB), and apparent diffusion coefficient map (ADC) which compose 3D bpMRI are each integral to prostate cancer detection~\cite{americancollegeofradiologyPIRADSProstateImaging2019}. Notably, no prior work has studied whether institution-specific synthetic 3D prostate bpMRI could improve inter-institutional domain adaptation. Examining this scenario could improve data scarcity mitigation, and consequently ML model performance within this space.

To address this gap, we propose CCELLA++: a novel LDM pipeline for simultaneously generating 3D prostate bpMRI conditioned on the full clinical report text. This pipeline operates with minimal additional data annotation beyond clinical routine, requiring only bpMRI and the radiologist report text (with PI-RADS score automatically extracted from the report text) for training. CCELLA++ builds atop the AxT2-only CCELLA pipeline from~\cite{grabkeLeveragingClinicalText2025} to produce richer synthetic output for improving ML model performance within the prostate cancer detection from bpMRI space. We hypothesized that this richer synthetic data would yield improved classifier performances beyond synthetic AxT2 alone, given the integral role of the HighB and ADC series in clinical prostate MRI interpretation~\cite{americancollegeofradiologyPIRADSProstateImaging2019}.

We then systematically investigate the use of synthetic prostate bpMRI for source-free domain adaptation--where only model weights but not training data are transferred between institutions. Specifically, we conducted a series of experiments to investigate the usability of synthetic data pretraining when fine-tuning on a data-limited, out-of-distribution dataset for predicting prostate cancer histopathology results from MRI. Two disjointed, artificial institutions were used: one larger silo sourced from one local institution (Institution 1), and one smaller silo sourced from two public, multi-center datasets (Institution 2). Our investigation used training data volumes reasonably accessible to a single medical institution. We investigated the performance of classifiers trained on progressively smaller real data volumes from Institution 2. We compared this to classifiers first pretrained with real or LDM-generated synthetic data from Institution 1 and then fine-tuned with Institution 2 data. To assess our CCELLA++ method and the performance contribution of synthetic bpMRI images, we conducted a parallel assessment using AxT2-only synthetic data generated by CCELLA~\cite{grabkeLeveragingClinicalText2025}. It was hypothesized that out-of-distribution pretraining would outperform training from scratch at smaller Institution 2 data volumes, with synthetic data performing comparably to real data for pretraining. It was also hypothesized that CCELLA++ pretraining would outperform CCELLA pretraining.

In summary, this paper makes the following contributions:
\begin{itemize}
	\item CCELLA++, a novel LDM pipeline for simultaneous 3D prostate bpMRI generation using full radiology report texts.
	\item Evidence that institution-specific synthetic bpMRI can outperform real data in data-limited source-free domain adaptation scenarios.
	\item Demonstration that bpMRI synthetic pretraining yields better generalizability over AxT2-only synthetic pretraining.
\end{itemize}

\section{Related Work}
\label{sec:relatedwork}

\subsection{Latent Diffusion Models for Medical Imaging}
Latent diffusion models (LDM) are a recent generative ML model architecture composed of a volume-compressing vector-quantized autoencoder (VQ-VAE) and a U-Net for latent image denoising~\cite{rombachHighResolutionImageSynthesis2022}. The VQ-VAE encoder first encodes an image to a fixed-distribution latent space. Gaussian noise is repeatedly added to the latent image over many (e.g. 1000) time steps. The U-Net is then trained to predict the noise added in each time step, enabling removal of the added noise (i.e. denoising). On inference, randomly generated Gaussian noise is iteratively denoised by the U-Net. The VQ-VAE decoder then decodes the denoised latent to yield the synthetic image.

Medical imaging literature has seen a recent surge in the number of LDM studies. Reported motivations for medical LDM use include augmenting datasets for rare pathologies and alleviating data annotation costs~\cite{kazerouniDiffusionModelsMedical2023}. LDM applications within this space include image generation, reconstruction, and segmentation~\cite{fanSurveyEmergingApplications2024}. While medical LDMs for MRI generation are typically uni-modal, recent literature has explored generating multiple MRI sequences for the same exam~\cite{kazerouniDiffusionModelsMedical2023}, particularly for brain MRI generation~\cite{jiangCoLaDiffConditionalLatent2023,xingCrossconditionedDiffusionModel2024,lupkePhysicsInformedLatentDiffusion2024}.

Grabke et al.~\cite{grabkeLeveragingClinicalText2025} is the most similar LDM work to our study, though with a few critical differences. Firstly, our study expanded the CCELLA-centric LDM framework to accommodate simultaneous generation of all bpMRI series for a single exam. We also investigated the effects of inter-institutional domain adaptation when pretraining a classifier on real and/or synthetic data. Finally, our work prioritized histopathology classification over PI-RADS classification during the classifier tasks, which is more clinically relevant for prostate cancer detection from MRI~\cite{sahaArtificialIntelligenceRadiologists2024}.

\subsection{Domain Shift and Adaptation}
Transfer learning occurs when learned knowledge from a pretrained ML model is used to improve the performance of a new model on a new task~\cite{panSurveyTransferLearning2010}. A primary motivation for transfer learning is to mitigate data scarcity, enabling improved performance on reduced training data. Domain shift is a common challenge to transfer learning, occurring when the target dataset is of different distribution than the source pretraining dataset~\cite{moreno-torresUnifyingViewDataset2012}. Domain shift sources include noise, label imbalance, and incomplete samples. Within medical imaging, domain shift can occur from differences in patient population, imaging machine properties and image acquisition protocols~\cite{guanDomainAdaptationMedical2022,mataArtificialIntelligenceAssisted2021}. Domain adaptation is a subset of transfer learning that seeks to mitigate domain shift.

Fine-tuning is a common domain adaptation method whereby a large-dataset-pretrained model is fine-tuned on a smaller target dataset~\cite{chenCloserLookFewshot2020,dhillonBaselineFewShotImage2020}. Both partial and entire model fine-tuning has yielded performance increases within recent medical imaging ML literature~\cite{motamedTransferLearningApproach2019,candemirTrainingStrategiesRadiology2021}.

Brugnara et al. is similar in concept to our study~\cite{brugnaraAddressingGeneralizabilityAI2024}. The authors of this study proposed the use of synthetic brain MRI from a generative adversarial network (GAN) to mitigate inter-institutional domain shift when training classifier models. The authors demonstrated that training classifier models on a combination of real and synthetic data improved zero-shot classifier performance on test data from an external, unseen institution~\cite{brugnaraAddressingGeneralizabilityAI2024}. In contrast, our work performed a more comprehensive domain adaptation experiment by training on multiple data subsets from the external institution, in addition to the zero-shot assessment. We additionally assessed the performances of models trained independently on real or synthetic data rather than only assessing their combination. Our study also used and improved upon an LDM framework instead of using a GAN for synthetic image generation.

\section{Methods}
\label{sec:methods}
\subsection{Dataset}
This retrospective study used biparametric MRI (bpMRI) of men at high-risk for prostate cancer. This study was approved by an institutional research ethics board with the need for informed consent waived. Two non-overlapping datasets were sourced for this study, forming the two simulated institutional datasets (Institutions). The ``Institution 1'' dataset consisted of 4152 images sourced from one healthcare institution. The ``Institution 2'' dataset consisted of 1658 images sourced from two publicly available datasets (1500 from the PI-CAI Grand Challenge~\cite{sahaArtificialIntelligenceRadiologists2024}, 158 from the Prostate158 dataset~\cite{adamsProstate158Expertannotated3T2022}). Institution 1 inclusion criteria were referral for prostate MRI, being treatment naïve, and MRI date between April 2009 and August 2021. Exclusion criteria were non-diagnostic image quality or incomplete exam. Institution 2 inclusion and exclusion criteria have been published~\cite{sahaArtificialIntelligenceRadiologists2024,adamsProstate158Expertannotated3T2022}.

Every bpMRI comprised: 1) the axial T2-weighted (AxT2) sequence, 2) diffusion weighted imaging at b$\ge$1000 s/mm\textsuperscript{2} (HighB), 3) apparent diffusion coefficient (ADC) maps. All HighB sequence b-values for Institution 1 were calculated at b=1600 s/mm\textsuperscript{2} whereas the Institution 2 HighB b-values varied. Institution 1 studies also included full radiology report texts as generated during clinical routine. All bpMRI images and reports were anonymized prior to this study. Institution 1 radiology classifications were reviewed and re-assessed as necessary to adhere to the latest prostate MRI guidelines~\cite{americancollegeofradiologyPIRADSProstateImaging2019} prior to this study~\cite{grabkeLeveragingClinicalText2025}. All exams had either PI-RADS score or histopathology result from within 6 months of the MRI. Consistent with the PI-CAI Grand Challenge, exams were assigned class labels (i.e. positive or negative) using histopathology results, with PI-RADS used only if histopathology information was missing~\cite{sahaArtificialIntelligenceRadiologists2024}. Images were preprocessed using the methods described in~\cite{grabkeLeveragingClinicalText2025}.

\begin{figure}
	\centering
	\includegraphics[width=1\linewidth]{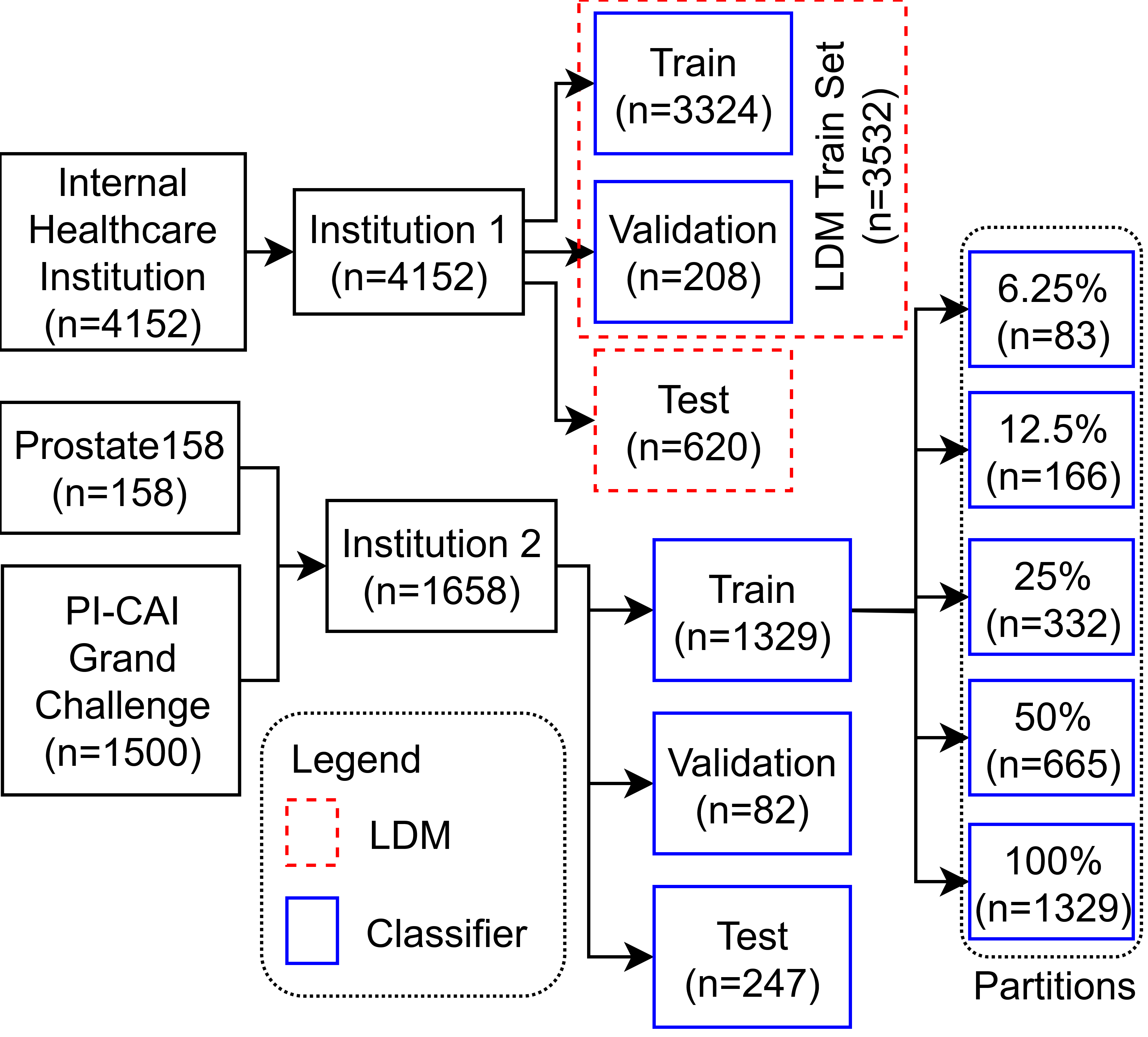}
	\caption{Overview of experimental data flow for both CCELLA++ and domain adaptation}
	\label{figdataflow}
\end{figure}

Each institutional dataset was randomly split into train, validation, and test sets targeting a ratio of 80:5:15. Splits were proportional by class within the respective Institution and occurred on a per-patient level to ensure all images from one patient were contained to a single split. Final train, validation, and test split sizes were 3324, 208, 620 for Institution 1 and 1329, 82, 247 for Institution 2 (respectively). Test sets were held out, unseen by all models until the end of the experiment when models were run once against these images to evaluate performance metrics. Figure \ref{figdataflow} depicts the data flow and partitions used in all parts of this study.

\subsection{Latent Diffusion Models}
The CCELLA-centric framework was used as the baseline LDM for this study and has already been validated against prior state-of-the-art~\cite{grabkeLeveragingClinicalText2025}. In brief, this framework used two pretrained, frozen components: 1) MAISI VQ-VAE~\cite{guoMAISIMedicalAI2025} for volume compression with a latent channel size of 4, 2) FLAN-T5 XXL LLM~\cite{chungScalingInstructionFinetunedLanguage2024} for report text encoding. The denoising U-Net and dual-head conditioning adapter were trained from scratch. The CCELLA LDM framework was trained on the AxT2 sequences from Institution 1.

CCELLA++ extended the CCELLA framework to enable bpMRI training and generation. We independently encoded and decoded each bpMRI sequence from a single exam, thus enabling reuse of the pretrained VQ-VAE despite its single-sequence restriction. We employed sequence-specific scaling factors to mitigate differences across sequence types. We then concatenated the three encoded latent images along the channel dimension to produce a latent channel size of 12, and modified the denoising U-Net architecture to accept 12 input and output channels. On inference, we split the denoised latent object every 4 channels and independently decoded each split to produce the different bpMRI images for a single exam. The LLM adapter was not modified during this process.

\textbf{Training details:} The CCELLA LDM was trained with identical hyperparameters as the original paper~\cite{grabkeLeveragingClinicalText2025}. The CCELLA++ LDM trained for 800 epochs with otherwise identical hyperparameters, with more epochs trained to accommodate for the additional HighB and ADC. Both LDMs were trained on a NVIDIA DGX workstation with n=4 Tesla V100 32GB VRAM GPUs. To maximize training data, both LDMs used a training set consisting of the combined train and validation splits from Institution 1 (n=3532) and trained to completion.

\textbf{Evaluation:} On inference, the LDMs each generated one synthetic exam for every real MRI exam in Institution 1. Images in the train and validation sets were used during the domain adaptation studies (see below). Images in the test set were used to assess LDM performance using the Kernel Inception Distance (KID) calculated in two ways~\cite{binkowskiDemystifyingMMDGANs2018} as well as the per-image Learned Perceptual Image Patch Similarity (LPIPS)~\cite{zhangUnreasonableEffectivenessDeep2018}.

KID was first evaluated in 3-dimensions with z-score normalized images using the Med3D ResNet50 pretrained on 23 datasets~\cite{chenMed3DTransferLearning2019}, and was the primary LDM performance metric for this study. KID was also evaluated in 2-dimensions with linear rescaling for each of the sagittal, coronal and axial planes using a ResNet50 pretrained on RadImageNet~\cite{meiRadImageNetOpenRadiologic2022}. Image rescaling methods were selected to match the training of each KID model. KID was chosen over the more widely used Fréchet Inception Distance (FID) metric~\cite{heuselGANsTrainedTwo2017} as KID is an unbiased estimate and exhibits increased reliability over FID when used to assess test sets of limited sample size~\cite{binkowskiDemystifyingMMDGANs2018}.

LPIPS compared synthetic perceptual quality distributions between the different LDM-generated sequences. LPIPS was calculated for each synthetic-real image pair in 3-dimensions using the same Med3D ResNet50 as the KID metric~\cite{chenMed3DTransferLearning2019}.

\textbf{Statistical Analysis:} A paired Wilcoxon signed-rank test was performed for each LPIPS distribution to directly compare AxT2 perceptual quality between CCELLA and CCELLA++ AxT2 sequences. A Friedman test followed by post-hoc Wilcoxon with Bonferroni correction was then used to compare between CCELLA++ generated sequences. This enabled insight into which sequences CCELLA++ had more difficulty perceptually reproducing.

\subsection{Domain Adaptation Studies}
To simulate domain adaptation at exponentially increasing levels of data restriction, five Institution 2 training dataset partitions were randomly selected. Partitions targeted 100\%, 50\%, 25\%, 12.5\% and 6.25\% of the total data volume for Institution 2. This resulted in 1329, 665, 332, 166, and 83 Institution 2 training cases (respectively). Partitions were proportional by class within the Institution 2 training set to ensure a consistent label distribution between partitions. Randomly selected exams within a single partition remained consistent across all models trained on that partition.

The domain adaptation study began with two classifiers trained from scratch: one on all Institution 1 real data and one on all Institution 1 synthetic data. These classifiers represented a zero-shot inference scenario. Three models were then trained for each Institution 2 data partition: one from scratch on the data partition; two fine-tuned from the pretrained Institution 1 classifiers, one with real data and another with synthetic, on the same Institution 2 data fraction.

All classifier models were trained in triplicate (n=3 trials) with average metrics reported to protect against training abnormalities, particularly among the low-data partitions. This domain adaptation study was performed once for each of the CCELLA and CCELLA++ LDMs using AxT2-only and bpMRI images (respectively). 

\textbf{Training details:} All classifiers used the EfficientNet-b0 architecture. Classifiers were trained with batch size 32, AdamW optimizer, and second-order polynomial learning rate scheduler on n=1 Tesla V100 32GB VRAM GPU. Classifiers trained from scratch used an initial learning rate of $10^{-4}$ targeting 16k training steps. This resulted in 155 epochs for Institution 1 and 385 epochs for Institution 2 when training from scratch. An initial learning rate of $10^{-5}$ for 385 epochs was used for all Institution 2 fine-tuning regardless of partition size. All images were clipped from 2\% to 98\%, z-score normalized, and zero-padded to $256 \times 256 \times 32$. Training data was augmented with random flips, rotations, translations and zooms. A validation loop occurred every 5 epochs, using highest validation average precision (AP) to determine the best model. The Institution 1 validation set was used during Institution 1 training (with real if training images were real, synthetic otherwise), and the Institution 2 validation set for Institution 2 training or fine-tuning. AP was selected over the area under the receiver operating characteristic curve (AUC) for checkpointing due to its higher robustness against label distribution changes. 

\textbf{Evaluation:} All classifiers were evaluated on Institution 2's test set with AP and AUC as primary metrics. Reported metrics were averaged between the three trials for each classifier. Primary comparisons were performed between pretraining methods for the same number of input sequences, with secondary comparisons performed between AxT2-only and bpMRI models for the same pretraining method.

\textbf{Statistical Analysis:} Similar to Bosma et al.~\cite{bosmaSemisupervisedLearningReportguided2023}, bootstrapping (n\_iterations = 1M, sampling with replacement) was used to calculate 95\% confidence intervals (CI). Two-sided p-values were calculated comparing Institution 1 synthetic pretraining to either Institution 1 real pretraining or Institution 2 training from scratch for every external partition. Two-sided p-values were calculated similarly to the CIs but with paired sampling, significance margin of 0.05, and Benjamini-Hochberg correction with false discovery rate controlled at 0.05.

\section{Results}
\label{sec:results}
\subsection{Latent Diffusion Model Performance}
\begin{table*}[t]
	\centering
	\caption{LDM Performance on Test Set. KID reporting mean $\pm$ std; LPIPS reporting median [IQR].}
	\label{table1}
	\renewcommand{\arraystretch}{1.2}
	\resizebox{\textwidth}{!}{
	\begin{tabular}{@{} ll rrrr r @{}}
		\toprule
		\textbf{LDM} & \textbf{Sequence}
		& \textbf{KID 3D}$\downarrow$
		& \textbf{KID 2D Axial}$\downarrow$
		& \textbf{KID 2D Coronal}$\downarrow$
		& \textbf{KID 2D Sagittal}$\downarrow$
		& \textbf{LPIPS 3D}$\downarrow$ \\
		\midrule
		CCELLA
		& AxT2
		& $0.0131 \pm 0.0017$
		& $-0.0078 \pm 0.0015$
		& $-0.0017 \pm 0.0016$
		& $-0.0018 \pm 0.0016$
		& $0.0025\ [0.0018,0.0037]$ \\
		\midrule
		\multirow{3}{*}{CCELLA++}
		& AxT2
		& $0.0128 \pm 0.0015$
		& $-0.0078 \pm 0.0016$
		& $-0.0021 \pm 0.0016$
		& $-0.0012 \pm 0.0017$
		& $0.0023\ [0.0017,0.0037]$ \\
		& HighB
		& $0.0138 \pm 0.0014$
		& $0.0214 \pm 0.0016$
		& $0.0199 \pm 0.0026$
		& $0.0184 \pm 0.0028$
		& $0.0053\ [0.0034,0.0088]$ \\
		& ADC
		& $0.0056 \pm 0.0020$
		& $0.0052 \pm 0.0016$
		& $-0.0017 \pm 0.0016$
		& $0.0000 \pm 0.0017$
		& $0.0029\ [0.0023,0.0040]$ \\
		\bottomrule
	\end{tabular}
}
\end{table*}

\begin{table*}[t]
	\centering
	\caption{Statistical Comparison of Sequences}
	\label{tablestats}
	\renewcommand{\arraystretch}{1.2}
	\resizebox{\textwidth}{!}{
	\begin{tabular}{@{} l l r r c r@{}}
		\toprule
		\textbf{Comparison} & \textbf{Test} & \textbf{Statistic ($W$)} & \textbf{$p$-value} & \textbf{$r_{\mathrm{rb}}$} & \textbf{LPIPS 3D diff.}\\
		\midrule
		CCELLA AxT2 vs CCELLA++ AxT2
		& Wilcoxon signed-rank
		& 90676 & 0.211 & $+0.058$ & $+0.0001\ [-0.0005,+0.0006]$\\
		CCELLA++ AxT2 vs CCELLA++ HighB
		& Wilcoxon (Bonf.\ $\alpha=0.0167$)
		& 16420 & \textbf{\textless0.001} & $-0.829$ & $-0.0026\ [-0.0054,-0.0008]$\\
		CCELLA++ AxT2 vs CCELLA++ ADC
		& Wilcoxon (Bonf.\ $\alpha=0.0167$)
		& 68232 & \textbf{\textless0.001} & $-0.291$ & $-0.0005\ [-0.0014,+0.0004]$\\
		CCELLA++ HighB vs CCELLA++ ADC
		& Wilcoxon (Bonf.\ $\alpha=0.0167$)
		& 25336 & \textbf{\textless0.001} & $+0.737$ & $+0.0020\ [+0.0002,+0.0052]$\\
		\bottomrule
	\end{tabular}
}
\end{table*}

Table~\ref{table1} describes the test set performance metrics for both the CCELLA and CCELLA++ LDMs using both sequence-level KID and per-image LPIPS. Friedman test resulted in $\chi^2(2)=421.6$ with $p<0.001$ for the CCELLA++ inter-sequence LPIPS measures, indicating significant difference among the three output sequences. LPIPS-derived statistical test results, including p-value and rank-biserial correlation ($r_rb$), can be found in Table~\ref{tablestats}. No statistically significant difference was found between LPIPS values for the CCELLA and CCELLA++ AxT2 sequences. All pairwise comparisons between CCELLA++ sequences were significantly different. An example synthetic exam for each of CCELLA and CCELLA++ can be found in the Supplementary Materials.

\subsection{Domain Adaptation}

\begin{figure*}
	\centering
	\includegraphics[width=\linewidth]{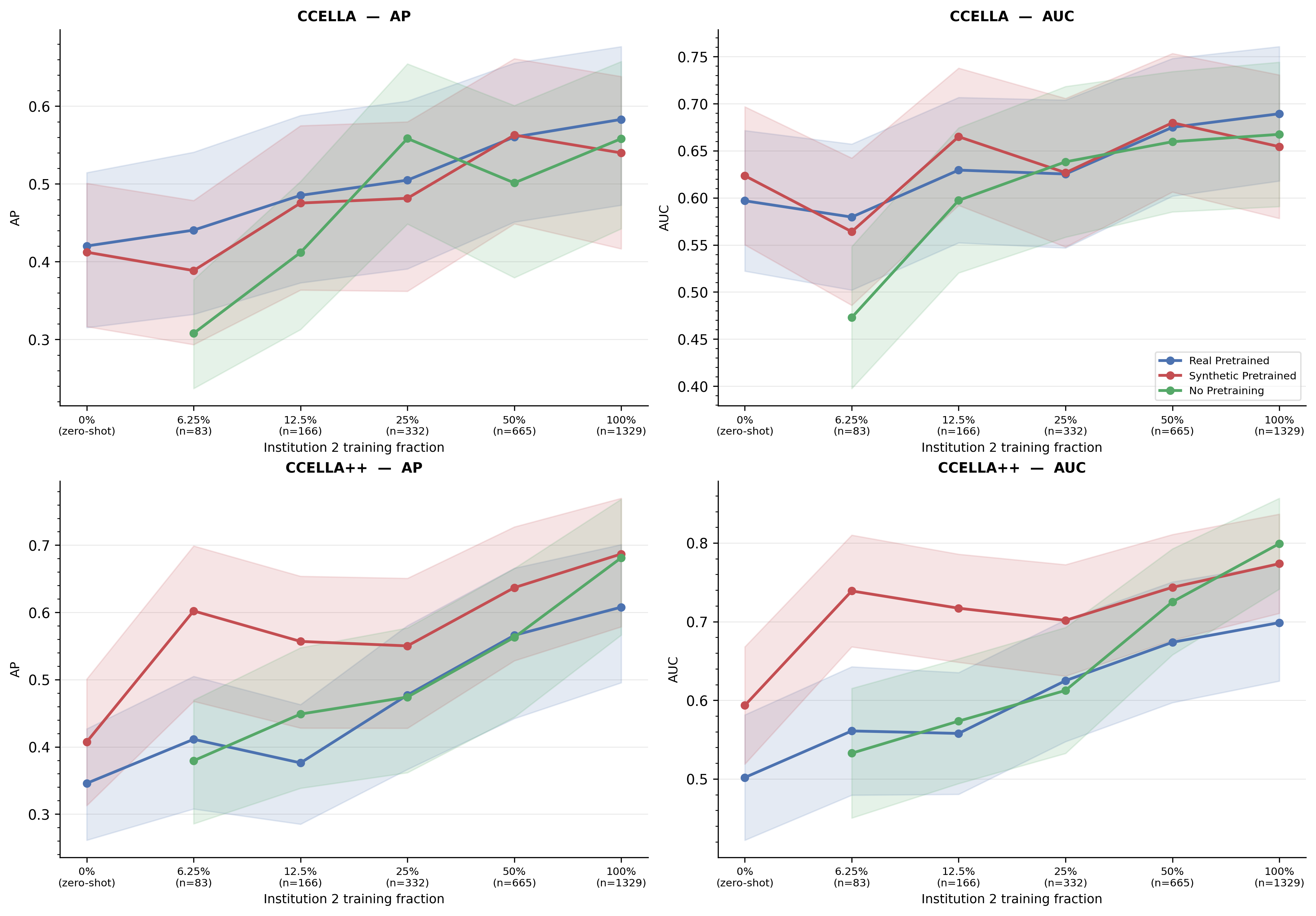}
	\caption{Classifier Average Precision (AP) and Area Under the Receiver Operating Characteristic Curve (AUC) Performances with 95\% Bootstrap Confidence Intervals (illustrated as painted areas, bootstrapping at 1M samples)}
	\label{figlines}
\end{figure*}

\begin{figure*}
	\centering
	\includegraphics[width=1\linewidth]{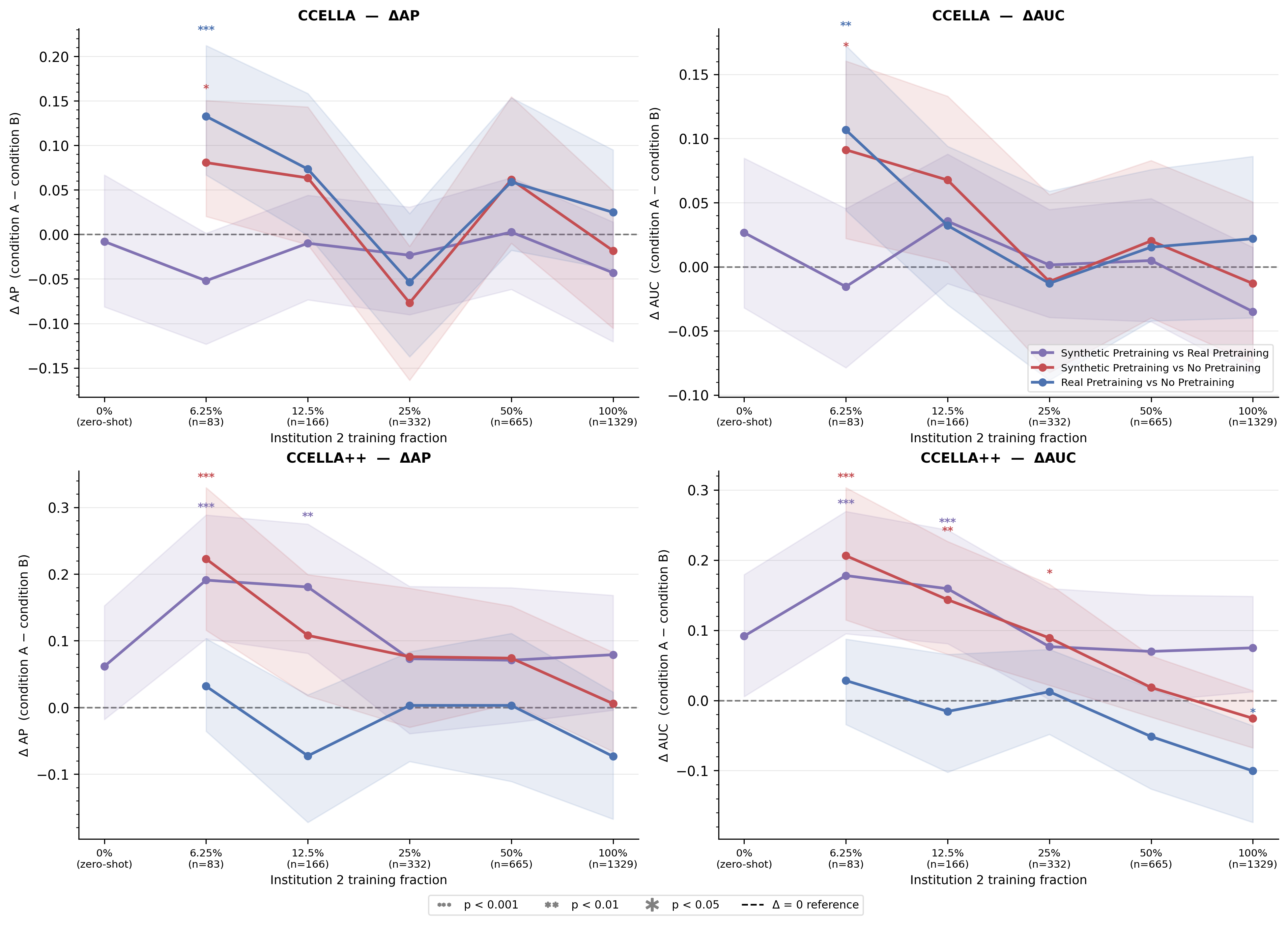}
	\caption{Differences in AP and AUC from Figure~\ref{figlines} within AxT2-only (top) and bpMRI (bottom) scenarios.}
	\label{fig3:fig}
\end{figure*}

\begin{figure*}
	\centering
	\includegraphics[width=1\linewidth]{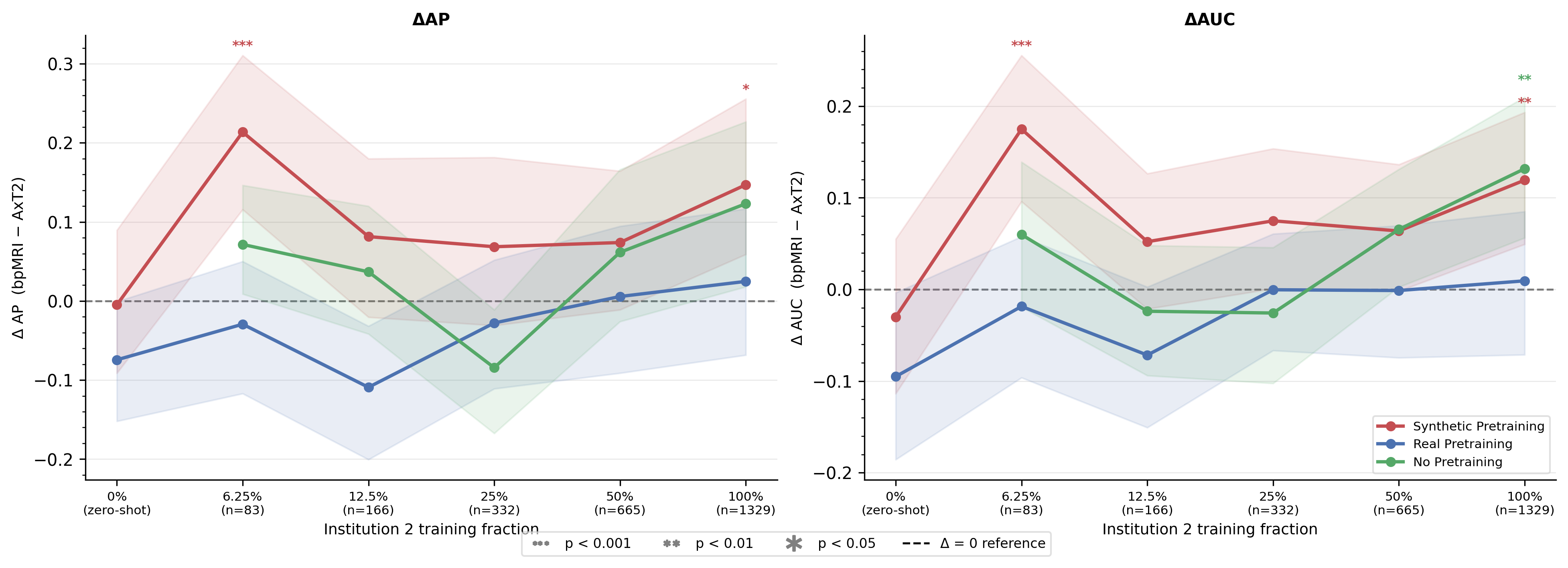}
	\caption{Differences in AP and AUC from Figure~\ref{figlines} for AxT2-only vs bpMRI Training.}
	\label{figdiffs_singlemulti}
\end{figure*}

Figure~\ref{figlines} illustrates the domain adaptation performances of all classifiers, with Figure~\ref{fig3:fig} illustrating scenario differences with statistical significance where applicable. Within the AxT2-only domain adaptation performances, pretraining was only found to significantly outperform training from scratch for the 6.25\% partition, regardless of pretraining method used. Synthetic data pretraining was not significantly different from real data pretraining at any partition.

Domain adaptation with bpMRI exhibited clearer trends than AxT2-only (see Figure~\ref{figlines} and Figure~\ref{fig3:fig}). Compared to training from scratch, real data pretraining did not differ significantly in performance at any partition except at 100\% where AUC performed significantly worse. Compared to real pretraining, synthetic pretraining exhibited significant increase in AP and AUC at the 6.25\% and 12.5\% partitions. Compared to training from scratch, synthetic data pretraining significantly outperformed training from scratch in AP and AUC at 6.25\% and in AUC through 25\%.

Results from comparing AxT2-only and bpMRI classifier performance are seen in Figure~\ref{figdiffs_singlemulti}. Classifiers trained from scratch on bpMRI exhibit significantly higher AUC compared to AxT2-only at full external data. Pretraining with synthetic bpMRI outperforms training on synthetic AxT2-only MRI with significance at both the 6.25\% and 100\% partitions. While pretraining with real bpMRI data underperforms AxT2-only in AP and AUC at lower data fractions, this difference was not statistically significant.

All domain adaptation numerical results, including p-values, can be found in Supplementary Materials.

\section{Discussion}
\label{sec:discussion}
CCELLA++ achieved comparable AxT2 3D KID with absence of statistically significant LPIPS difference (p=0.211) to the CCELLA AxT2. This validates that the addition of the HighB and ADC modalities did not impede CCELLA++ AxT2 generative performance compared to CCELLA. CCELLA++ exhibited significantly different LPIPS metrics ($p<0.001$) for every sequence comparison despite overlapping LPIPS IQRs particularly between the AxT2 and ADC. This is likely because the Wilcoxon signed-rank test operates on paired differences; thus even though their marginal distributions overlap, the sequences still differ consistently in LPIPS metric for each real-synthetic image pair. Sequence LPIPS comparisons against HighB exhibited rank-biserial correlations and median differences suggesting that CCELLA++ struggled the most with this sequence (Table~\ref{tablestats}). Recent literature on LDM use within the prostate MRI space suggests that the varying b-values can play a role in domain shift~\cite{liDeepLearningBased2024}. To that end, the variable b-values for the images in this study could have played a role in the worse LPIPS performance.

This study performed domain adaptation experiments for both CCELLA and CCELLA++. The CCELLA experiment only demonstrated pretraining (real or synthetic) value for both AP and AUC at the 6.25\% partition size (Figure~\ref{fig3:fig}). Real versus synthetic Institution 1 pretraining did not demonstrate a clear performance difference. 

In contrast, the CCELLA++ domain adaptation experiment depicted a clearer trend where synthetic data repeatedly outperformed real data pretraining and pretraining from scratch at lower data volumes (Figure~\ref{fig3:fig}). Perhaps the most interesting statistic is that real pretraining AUC performed significantly worse than training from scratch at 100\% Institution 2 data, whereas synthetic data pretraining did not perform significantly differently. This reaffirms that there exists domain shift between these datasets. These results extend~\cite{brugnaraAddressingGeneralizabilityAI2024} by demonstrating that the generalizability advantages of synthetic pretraining are maintained in both bpMRI and source-free domain adaptation scenarios.

When comparing AP and AUC across bpMRI and AxT2 scenarios, a few trends can be noted (Figure~\ref{figdiffs_singlemulti}). Among classifiers trained from scratch, bpMRI outperforming AxT2-only in AUC at full external data volume is consistent with clinical prostate reporting guidelines which place emphasis on the HighB and ADC~\cite{americancollegeofradiologyPIRADSProstateImaging2019}. Classifiers trained with CCELLA++ synthetic images maintain this trend as hypothesized, with higher numerical performance compared to CCELLA-pretraining at almost every partition. Interestingly, this trend was not maintained with real data pretraining. Instead, real pretrained bpMRI classifiers performed numerically worse than their AxT2-only counterparts at lower data fractions, though without statistical significance. This is likely due to domain shift between the two datasets, particularly with respect to the HighB images having variable b-values. To that end, the poorer domain-adaptation performance of the real pretrained bpMRI classifier and the significantly higher performance of the CCELLA++ pretrained classifier compared to real pretraining at lower data fractions affirms that CCELLA++ generated synthetic data can improve model generalizability in data-scarce transfer learning scenarios.

This study presented a few limitations and opportunities for future work. Firstly, this study did not directly investigate individual data sample quality. Sample quality variance likely contributed to the variances in classifier performance at smaller data partitions despite efforts to mitigate this through metric averaging. It would be worthwhile to investigate methods for quantifying individual real and synthetic medical image quality. Secondly, this study explored latent concatenation with different per-sequence scaling factors for generating bpMRI in order to reuse the pretrained VQ-VAE. However, CCELLA++ exhibited worse HighB performance compared to the AxT2 and ADC. This presents an opportunity for future work to further balance intra-bpMRI training for CCELLA++. Finally, this study only explored text- and class- conditioning as per its predecessor model. This presents an opportunity to further improve CCELLA++ performance by incorporating additional LDM conditioning information such as biopsy data or anatomical segmentations.

\section{Conclusion}
\label{sec:conclusion}
In summary, this study presents CCELLA++ for synthetic prostate bpMRI generation as well as a domain adaptation experiment investigating the utility of synthetic data for pretraining before fine-tuning on out-of-distribution data. We successfully demonstrated that our method generates synthetic bpMRI while maintaining similar AxT2 image quality to its predecessor. We also demonstrated that at low external data volumes, pretraining with our method's synthetic bpMRI data outperforms both pretraining with real bpMRI data and training from scratch at the second institution. Finally, we demonstrated that pretraining with synthetic bpMRI can outperform synthetic AxT2-only pretraining, validating the downstream utility of our method over its predecessor.

These results suggest that source-free domain adaptation after pretraining on synthetic data generated by our method can result in improved performance compared to real data pretraining and training from scratch on small out-of-distribution datasets. Our findings of synthetic data yielding greater out-of-distribution generalizability are consistent with existing literature~\cite{brugnaraAddressingGeneralizabilityAI2024}. Future work should explore methods for quantifying medical image sample quality, balancing bpMRI LDM training, and conditioning the LDM with additional information.

\bibliographystyle{IEEEtran}
\bibliography{bstctl,library}

\end{document}